\documentclass[twocolumn,twoside,slac_two]{revtex4}
\usepackage{graphicx}
\usepackage{fancyhdr}
\begin{document}

\title{Some consequences for LHC experiments from the results of cosmic ray investigations above the knee}

\author{A.A. Petrukhin}
\affiliation{National Research Nuclear University MEPhI, Moscow 115409, Russia}

\begin{abstract}
Experimental data obtained in cosmic ray investigations at energies above the knee are discussed. All features observed at these energies starting from the knee, changes of mass composition and concluding with various types of unusual events are analysed in the framework of the new hadron interaction model  producing blobs of quark gluon matter. Consequences for LHC experiments are considered. 
\end{abstract}

\maketitle
\thispagestyle{fancy}

\section{INTRODUCTION}
Forthcoming experiments at the LHC in the interval 2$\div$14~TeV for p-p interactions sharply increase the interest in various predictions (theoretical and experimental) for this energy region.

It is known that many physicists predict the appearance of new particles and states of matter in the TeV energy region, but none of them predicts the appearance of these new objects with large cross section. Values of usually predicted cross-sections are evaluated in nb or even pb. Of course processes with such small cross sections cannot be observed in cosmic rays.

In cosmic ray experiments, the interval 2$\div$14~TeV in the center of mass system corresponds to \mbox{$2\times{10}^{15}{\div}10^{17}$~eV} in laboratory system, but namely in this energy region the following phenomena were observed: the knee in energy spectrum, change of mass composition and appearance of various unusual events. For their observations in cosmic ray experiments a rather large cross section for new processes is required. If these results obtained in cosmic rays are really physical (not methodical or erroneous) then corresponding phenomena can be found in LHC experiments.

\section{RESULTS OF CR EXPERIMENTS AT LHC ENERGIES}
These results can be separated into two basic groups:
\begin{itemize}
\item unusual events, which cannot be explained in the framework of existing theoretical models and approaches;
\item changes of measured EAS parameters as a function of energy, which can be interpreted as changes in the interaction model.
\end{itemize}

Various types of unusual events were detected in hadron experiments at high altitudes and in muon investigations. They were discussed in many papers (see f.e. \cite{ref1,ref2}) therefore in this paper we only list them. In hadron experiments these are: halos, alignment, penetrating cascades, long-flying component, Centauros and Anti-Centauros; in muon experiments: excess of muons in their energy spectrum at energies more than several tens TeV, observation of VHE ($>$~100~TeV) muons in various experiments. Of course each of these events can be explained by large physical or statistical fluctuations, but it is practically impossible to explain in this way the full set of these unusual events and phenomena. In addition, and this is very impotent, practically all these unusual events are observed at energies above the knee. Attempts to find similar events in accelerator experiments at beam energies which correspond to cosmic ray energies below the knee did not give positive results. Therefore it is possible to consider these events as evidence of new physics at energies above the knee.
	
If it is believed that the hadron interaction model is not changed in the TeV energy region, all changes in EAS parameter measurements versus primary particle energy can be explained by changes of the cosmic ray energy spectrum and/or mass composition (cosmophysical approach). For that, existing models of acceleration of particles in the Galaxy, and keeping them within it, predict that at knee energies protons begin to reach their acceleration energy limit and leave the Galaxy. Then, with the increase of energy per nucleus, helium nuclei begin to leave the Galaxy, etc.

\begin{figure}[ht]
\centering
\includegraphics[scale=0.75]{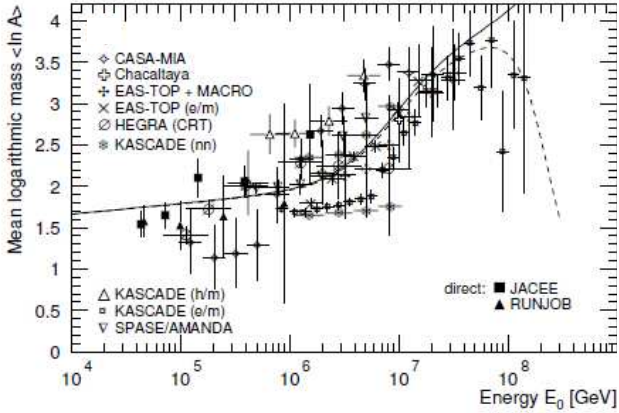}
\caption{Results of mean logarithmic mass evaluations from $N_\mu/N_e$-ratio measurements \cite{ref3}.} 
\label{fig01}
\end{figure}

\begin{figure}[ht]
\centering
\includegraphics[width=80mm]{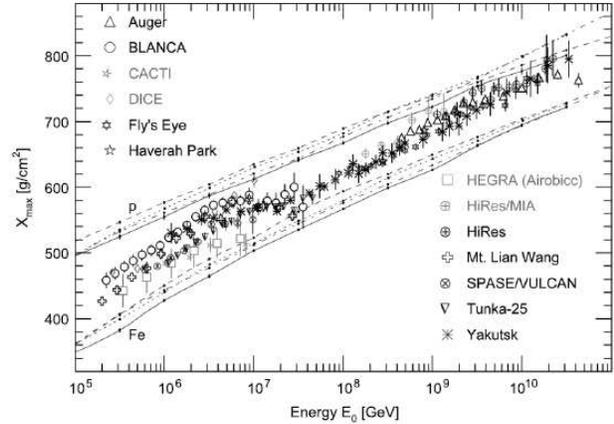}
\caption{Results of $X_{\rm max}$ measurements \cite{ref4}.} 
\label{fig02}
\end{figure}

In the framework of these cosmophysical models the mass composition of cosmic rays must be changed in favour of heavy nuclei, first slowly then more quickly up to iron nuclei. But the experimental situation is the opposite. Evaluations of the average logarithm of mass number $\langle\ln{A}\rangle$ based on $N_\mu/N_e$-ratio (Figure~\ref{fig01} \cite{ref3}) and $X_{\rm max}$ (Figure~\ref{fig02} \cite{ref4}) measurements give first a relatively sharp and strong growth of this value and then a slow decrease down to protons. In this situation consideration of an alternative explanation of the behaviour of the measured EAS parameters (nuclear-physical approach) is rather substantiated.

In the nuclear-physical approach, it is assumed that the primary cosmic ray energy spectrum and mass composition above the knee are not  seriously changed. All changes in the measured EAS parameters (the knee in EAS size spectrum, the increase of the $N_\mu/N_e$-ratio, the decreasing of $X_{\rm max}$ elongation rate, etc) are produced by the inclusion of new physical processes. 

\section{HADRON INTERACTION MODEL WITH QGM PRODUCTION}

To explain all unusual events and phenomena from a single point of view a new  interaction model  must satisfy the following requirements: threshold behaviour of cross section (unusual events begin to appear at energies more than several PeV), large value of cross-section (to be observed in cosmic ray experiments), large orbital momentum (to explain alignment), large yield of leptons (to explain VHE muon excess) and widening possibilities of EAS development (to explain changes in measurements of $N_e$, $N_\mu/N_e$, $X_{\rm max}$ and other parameters).
	
The production of quark-gluon plasma (better to speak about quark-gluon matter -- QGM) provides the fulfilment of conditions listed above: threshold behaviour, since high temperature (energy) for its appearance is required; large cross section, since in this case transition from quark-quark interaction to the interaction of many quarks and gluons occurs and the geometrical cross section changes from $\pi\lambda^2$ to $\pi{R}^2$, where $R$ is the size of the QGM blob which is not less than the nucleon radius; large orbital momentum in peripheral ion-ion collisions, since, as was shown in \cite{ref5}, the quark-gluon plasma as a globally polarized object is produced with orbital momentum, $L$, which is proportional to $\sqrt{s}$, and correspondingly a large centrifugal barrier appears.

The two last points are very important for explaining various phenomena observed in cosmic rays. The transition to collective interactions of many quarks drastically changes the energy in the center of mass system 
\begin{equation}
\sqrt{s}=\sqrt{2m_{\rm N}E_1} \rightarrow \sqrt{2m_{\rm c}E_1},
\end{equation} 
where $m_{\rm N}$ is nucleon mass, $m_{\rm c}$ is compound mass of many interacting quarks in nuclei of nitrogen or oxygen ($m_{\rm c}{\geq}m_{\rm N}$). A sharp change of $\sqrt{s}$ leads to a corresponding change of the value of orbital momentum, $L$, and the centrifugal barrier 
\begin{equation}
V(L)=L^2/(2mr^2).
\end{equation} 

This barrier will be high for light particles ($u$, $d$, $s$, $c$, $b$-quarks) but low for heavy particles ($t$-quarks). The production of  $t \bar t$-quark pairs with large mass changes the EAS  development. 
Decays of top-quarks $t(\bar t)\rightarrow{W}^+(W^-)+b(\bar b)$ and consequent decays of $W$-bosons and $b$-quarks give an excess of very high energy muons and neutrinos and also multiple hadrons with large transferred momenta. 

The possibility of explaining various unusual events in the framework of the new model was discussed in papers \cite{ref2,ref6}. The role of VHE muons in  explaining the penetrating cascades observed in the Pamir experiment \cite{ref7} was discussed in \cite{ref2} and in more detail in \cite{ref8}. As a whole a model of QGM production can explain all unusual events observed in cosmic rays.

\section{DESCRIPTION OF THE COSMIC RAY ENERGY SPECTRUM IN THE FRAMEWORK OF THE QGM MODEL}

In the framework of this model, hadrons will be produced at higher altitudes in the atmosphere than usual. Decays of $W$-bosons into leptons and hadrons give very large fluctuations in EAS development, change the average transition curve and can lead to an underestimation of the EAS energy and appearance of other phenomena: young and old showers, large transferred momenta, changes of $X_{\rm max}$ and $N_\mu/N_e$-ratio, etc. To evaluate the EAS energy it is necessary to take into account the energy carried away by top-quarks. Of course some part of this energy will be re-injected into EAS development. But in a first approximation, to simplify consideration it is possible to assume that all the energy, $\varepsilon_{\rm t}$, carried away by top-quarks is missing.
Then the energy in the centre of mass system is decreased ($\sqrt{s}-\varepsilon_{\rm t}$, where $\varepsilon_{\rm t}>2m_{\rm t}$) and correspondingly the EAS energy will be less than the energy of the primary particle
\begin{equation}
E_{\rm EAS}=\frac{(\sqrt{2m_{\rm c}E_1}-\varepsilon_{\rm t})^2}{2m_{\rm c}}.
\end{equation}
By this reason we obtain the steepening of the observed spectrum. 

Since for the production of quark-gluon matter, not only is high temperature (energy) required but also high density, it is reasonable to assume that at first QGM will be produced in nucleus-nucleus interactions but not in proton-nucleus interactions. This means that in cosmic rays the first component (at the same energy per nucleus) which will interact with  the production of QGM will be iron nuclei (or heavier ones), then more lighter nuclei, and finally the last ones protons. 

\begin{figure}[!t]
\centering
\includegraphics[width=80mm]{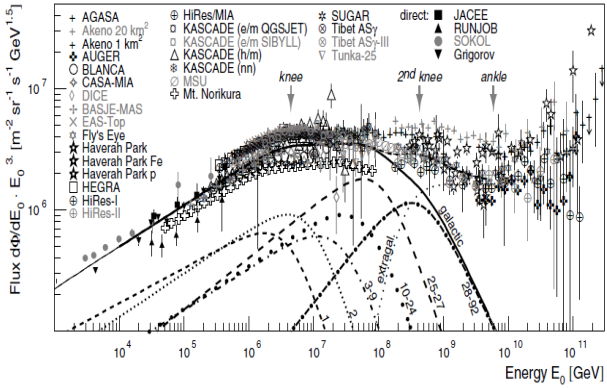}
\caption{The all-particle energy spectrum of primary cosmic rays and its explanation in the framework of the cosmophysical approach \cite{ref9}.} 
\label{fig03}
\end{figure}

\begin{figure}[ht]
\centering
\includegraphics[width=80mm]{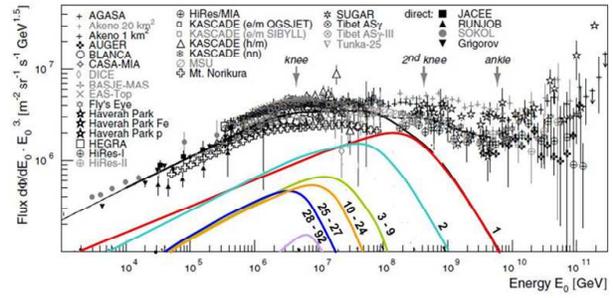}
\caption{New version of the contribution of various cosmic ray components in the measured energy spectrum.} 
\label{fig04}
\end{figure}

The difference between cosmophysical and nuclear-physical approaches to the interpretation of results of EAS measurements is demonstrated in  Figure~\ref{fig03} and Figure~\ref{fig04}. The Figure~\ref{fig03} from \cite{ref9} illustrates the changes of the contribution of various cosmic ray components into the all-particle energy spectrum in the framework of the cosmophysical approach. Figure~\ref{fig04} illustrates the contribution of various components in the same spectrum in the framework of the nuclear-physical approach. 

Figure~\ref{fig04} is constructed by the analogy of Figure~\ref{fig03}. The curves in this figure are not a result of calculations; they only illustrate the principal change of the approach. In Figure~\ref{fig03}, {\it real changes} of primary spectra of different cosmic ray species are shown. In Figure~\ref{fig04}, {\it the results of measurements} of different cosmic ray species are shown, which are really  not related with changes of their spectrum slope.

\begin{figure}[ht]
\centering
\includegraphics[angle=-90, width=80mm]{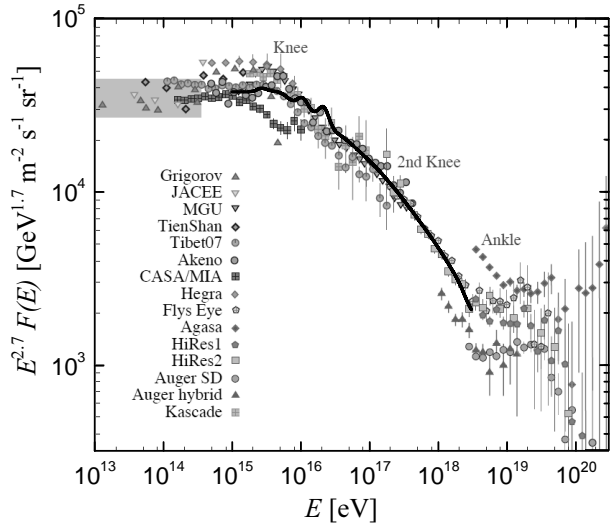}
\caption{Calculated spectrum and experimental data.} 
\label{fig05}
\end{figure} 

So, in this approach the knee is the result of our interpretation of observational data.

For more detailed comparison of predictions of the QGM model with experimental data, corresponding calculations were performed. The key question for this model is a value of threshold energy at which QGM blobs begin to be produced and their mass for various nuclei. Good results were obtained with the following dependences 
\begin{equation}
E_{\rm th}= E_{\rm knee} \left(56/A\right)^{0.5},
\end{equation}
\begin{equation}
m_{\rm c}=2m_{\rm N}A^{0.25}.
\end{equation}

Comparison with experimental data is shown in Figure~\ref{fig05}.

The energy spectrum obtained in the framework of the simplest model surprisingly well describes the experimental data. Observed changes of composition are also explained. A sharp increase of average mass above the knee is the result of additional detection of EAS from heavy nuclei. At higher energies a slow transition to more light nuclei (up to protons) begins. Results of composition changes obtained by means of $N_\mu$ measurements are also explained. The number of muons is increasing as a result of muon production not only in the usual scheme of EAS development, but also through decays of heavy particles. 

\section{CONSEQUENCES FOR LHC EXPERIMENTS}
If the considered nuclear-physical approach is correct then the following predictions can be made for LHC experiments directed in the search for new physics.

Firstly, for searches of quark-gluon matter with the described properties it is necessary to use nucleus-nucleus collisions. For pp-interactions the threshold energy can be more than $10^{17}$~eV for cosmic rays and more than 14~TeV for LHC (unfortunately in cosmic ray experiments it is impossible to check this circumstance). To correspond to conditions of cosmic ray experiments it is better to use nitrogen-nitrogen or oxygen-oxygen interactions. 

Secondly, the most clear signature to verify the validity of the considered approach in LHC experiments would be a sharp increase of top-quark production. For that, existing procedures of the search of top-antitop pairs -- ``lepton-jet'', ``multi-jet'' and ``dilepton'' -- can be used. But it is necessary to remark that usually production of $t\bar t$-quarks in pp-interactions is considered. In nucleus-nucleus interactions the total number of secondary particles and jets will be much greater, and it is possible that the development of new methods of data analysis will be required.

Another possibility is to search for missing energy which will be taken away by three types of neutrinos. Evaluations show that its value can be about 15\% of the beam energy.

A very good signature in the search for new physics is the behaviour of the considered characteristics as a function of the energy and mass of beam particles. It is important that for heavy nuclei the threshold energy must be less than for light nuclei.

\begin{figure}[!t]
\centering
\includegraphics[width=80mm]{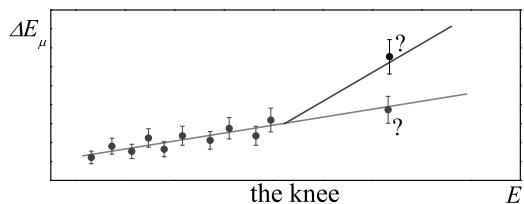}
\caption{Possible behaviour of the EAS muon energy deposit above the knee.} 
\label{fig06}
\end{figure} 

In this case the situation is similar to experiments in cosmic rays, in which the energy deposit of EAS muon component below and above the knee will be measured (Figure~\ref{fig06}). If the new state of matter appears at the knee energy the energy deposit and also behaviour of other EAS characteristics must be changed.  

\bigskip
\section{CONCLUSION}
Of course the considered model of new physics connected with QGM blob production cannot be the only 
one and other possibilities exist. But the need to obtain a large cross-section and to suppress decays of new heavy particles (or state of matter) into light quarks makes a choice in favour of QGM very alluring.

\bigskip 
\begin{acknowledgments}
The author thanks Alexey Bogdanov and Rostislav Kokoulin for big help in the fulfilment of this work. The work was supported by Ministry of Education and Science of the Russian Federation and RFBR (grant 09-02-12270-ofi-m).
\end{acknowledgments}


\bigskip 
\begin{thebibliography}{9}   

\bibitem{ref1}
S.A. Slavatinsky, Nucl. Phys. B (Proc. Suppl.) 122 (2003) 3.
\bibitem{ref2}
A.A. Petrukhin, in Proceeding of the Vulcano Workshop "Frontier Objects in Astrophysics and Particle Physics", 2004, Ed. F. Giovanelli \& G. Mannocchi, 489.
\bibitem{ref3}
J. H$\ddot{\rm o}$randel, Int. J. Mod. Phys. A 20 (2005) 6753.
\bibitem{ref4}
J. Bl$\ddot{\rm u}$mer et al. Progress in Part. and Nucl. Phys. 63 (2009) 293.
\bibitem{ref5}
Zuo-Tang Liang and Xin-Nian Wang: Phys. Rev. Lett. 94 (2005) 102301.
\bibitem{ref6}
A.A. Petrukhin, Nucl. Phys. B (Proc. Suppl.) 175-–176 (2008) 125.
\bibitem{ref7}
T. Arisava et al., Nucl. Phys. B. 424 (1994) 241.
\bibitem{ref8}
A.A. Petrukhin et al., Nucl. Phys. B (Proc. Suppl.) 196 (2009) 165.
\bibitem{ref9}
J. H$\ddot{\rm o}$randel, Mod. Phys. Lett. A22 (2007) 1533.


\end{thebibliography}
\end{document}